\documentclass[
preprint,
superscriptaddress,
amsmath,amssymb,
showkeys,
longbibliography,
aps,
prb,
]{revtex4-2}

\usepackage{threeparttable}
\usepackage{setspace}
\usepackage{makecell}
\usepackage{amsmath}
\usepackage{graphicx}
\usepackage{dcolumn}
\usepackage{bm}
\usepackage{hyperref}
\usepackage[mathlines]{lineno}
\usepackage{textcomp}
\usepackage{gensymb}
\usepackage{subfigure}
\usepackage{url}
\usepackage{xcolor}
\usepackage{hhline}
\usepackage[version=3]{mhchem}
\usepackage{multirow}
\usepackage[american]{babel}
\usepackage{textgreek}
\usepackage{miller}

\definecolor{red}{rgb}{0.6,0.1,0.1}

\begin{document}

\title{Anisotropic Core-Shell Swift Heavy Ion Tracks in \textit{\textbeta}-\ce{Ga2O3}}

\author{Huan He} 
\affiliation{School of Nuclear Science and Technology, Xi'an Jiaotong University, Xi'an, Shaanxi 710049, China}

\author{Jiayu Liang} 
\affiliation{School of Nuclear Science and Technology, Xi'an Jiaotong University, Xi'an, Shaanxi 710049, China}

\author{Shaowei He} 
\affiliation{School of Nuclear Science and Technology, Xi'an Jiaotong University, Xi'an, Shaanxi 710049, China}

\author{Jiahui Zhang} 
\affiliation{Materials Science and Environmental Engineering Unit, Tampere University, Tampere, 33720, Finland}

\author{Ziqi Cai}  
\affiliation{Institute of Modern Physics, Chinese Academy of Sciences, Lanzhou 730000, China}

\author{Tan Shi} 
\affiliation{School of Nuclear Science and Technology, Xi'an Jiaotong University, Xi'an, Shaanxi 710049, China}

\author{Yanwen Zhang}  
\affiliation{Department of Nuclear Physics, China Institute of Atomic Energy, Beijing, 102413, China}

\author{Flyura Djurabekova} 
\affiliation{Department of Physics and Helsinki Institute of Physics, University of Helsinki, P.O. Box 43, FI-00014, Finland}

\author{Hang Zang} \email{zanghang@xjtu.edu.cn}
\affiliation{School of Nuclear Science and Technology, Xi'an Jiaotong University, Xi'an, Shaanxi 710049, China}

\author{Chaohui He} \email{hechaohui@xjtu.edu.cn}
\affiliation{School of Nuclear Science and Technology, Xi'an Jiaotong University, Xi'an, Shaanxi 710049, China}

\author{Junlei Zhao} \email{junlei.zhao@mrdi.org.hk}
\affiliation{The Hong Kong Microelectronics Research and Development Institute, Hong Kong, 999077, China}


\begin{abstract}

Swift heavy ion (SHI) irradiation generates nanoscale ion tracks through intense electronic excitation, yet the microscopic mechanisms governing their morphology and phase stability in low-symmetry oxides remain poorly understood. 
Here, a multiscale atomistic simulation framework is used to investigate the formation and recovery of SHI-induced tracks in monoclinic $\beta$-\ce{Ga2O3} over a wide range of electronic energy losses ($S_\mathrm{e}$) and crystallographic orientations. 
A sequence of distinct structural responses is identified with increasing $S_\mathrm{e}$: (\textit{i}) complete lattice recovery at low $S_\mathrm{e}$; (\textit{ii}) recrystallization into a metastable $\gamma$-\ce{Ga2O3} phase at intermediate $S_\mathrm{e}$; and (\textit{iii}) the formation of core-shell ion tracks at high $S_\mathrm{e}$, consisting of an amorphous core surrounded by a recrystallized $\gamma$-phase shell. 
Despite the essentially isotropic initial energy deposition, the final ion-track morphology exhibits pronounced crystallographic anisotropy, governed by orientation-dependent recovery dynamics. 
The superior recrystallization along the \hkl[010] direction is attributed to its exceptionally high elastic stiffness. 
Notably, SHI irradiation perpendicular to the \hkl(100) plane induces a more severe structural response at low $S_\mathrm{e}$ ($\leq 10$~keV/nm), however, at higher $S_\mathrm{e}$, it yields a smaller residual ion track compared to the other orientations. 
The simulated ion-track sizes show excellent quantitative agreement with the available experimental measurements over a wide range of $S_\mathrm{e}$ values.
These findings establish a unified atomic-scale picture of core–shell track formation and anisotropic recovery in $\beta$-\ce{Ga2O3}.


\end{abstract}

\maketitle
\newpage

Swift heavy ion (SHI) irradiation, defined as the bombardment of materials with massive ions at ultrahigh kinetic energies ($E_\mathrm{k} \geq 1$~MeV/amu), has served as a powerful tool for probing fundamental radiation damage mechanisms and deliberately tailoring material properties~\cite{ridgway2011role, ochedowski2014graphitic, zhang2015ionization, lang2020fundamental, medvedev2023frontiers, amekura2024latent}.
The SHI energy is deposited primarily through electronic energy loss ($S_\mathrm{e}$) via intense ionization of the target's electron subsystem and triggers a cascade of physical processes: electron-electron scattering, non-equilibrium electron-lattice coupling, and subsequent lattice heating. 
Depending on the magnitude of $S_\mathrm{e}$ and the properties of the material, this can lead to irreversible lattice transformations, including ion-track formation~\cite{ridgway2013tracks, sequeira2021unravelling, amekura2024latent, hu2025latent}, phase transition~\cite{abdullaev2025ions, gupta2025unveiling}, elemental segregation~\cite{ochedowski2014graphitic}. 
The ion tracks produced by SHI are conventionally described as cylindrical damage regions with radial symmetry, reflecting the assumed isotropic dissipation of electronic excitation around the ion trajectory. 
This description is well supported by track measurements normal to high-symmetry crystal planes~\cite{wierbik2026anisotropic} and in amorphous matrices~\cite{kluth2008fine, ridgway2013tracks, bierschenk2023formation, ge2026ion}, where rotational invariance justifies cylindrical track cores and underpins thermal-spike-based models. 
However, whether such radial symmetry remains valid in low-symmetry crystals has only recently been critically examined~\cite{wierbik2026anisotropic}.

The monoclinic $\beta$-\ce{Ga2O3}, an emerging ultrawide-bandgap semiconductor for power, optical, and radiation-hardened electronics, provides a stringent test of this assumption~\cite{pearton2018review, SEE-Ga2O3-ai2021degradation, pearton2025perspective}. 
In recent years, significant research efforts have been devoted to studying the response to SHI irradiation of \ce{Ga2O3}~\cite{SHI_Ga2O3_manikanthababu2020swift, SEE-Ga2O3-ai2021degradation, SEE-Ga2O3-zhu2024swift, SEEGa2O3-ma2025single, ion_tracy2016anisotropic}. 
However, the majority of the experimental work has focused on single event effects and macroscale electrical degradation at the device level~\cite{SEEGa2O3-ma2025single, SEE-Ga2O3-ai2021degradation, SEE-Ga2O3-zhu2024swift}. 
At the material level, several experimental studies~\cite{ion_tracy2016anisotropic, ai2019radiation, iancu2025intrinsic} have provided valuable insight into the formation of the ion-track at low $S_\mathrm{e}$. 
However, the atomistic mechanisms elucidated by atomic-scale computational modeling remain rather limited due to the lack of suitable interatomic potentials (IAPs).

Recently, molecular dynamics (MD) simulations driven by machine learning (ML) IAPs have also been effectively used to investigate phase transformations induced by low-$S_\mathrm{e}$ ion-irradiation in a complex \ce{Ga2O3} system~\cite{zhao2023complex}, serving as an essential complement to experimental studies~\cite{azarov2023, zhao2025crystallization, SFazarov2025phase, abdullaev2025ions}. 
However, the underlying mechanisms in these cases are governed almost exclusively by nuclear stopping power ($S_\mathrm{n}$), resulting in ballistic collision cascades in which atoms are displaced through direct binary knock-on events.
Therefore, it remains unclear whether comparable structural transitions can be activated under high-$S_\mathrm{e}$ SHI irradiation, where primary damage is introduced via ultrafast electronic excitation, columnar thermal spikes, and subsequent lattice destabilization. 

In this work, we present a multiscale computational modeling of SHI-induced track formation in $\beta$-\ce{Ga2O3}, combining electronic energy deposition and evolution with ML-MD simulations to capture the full evolution from initial excitation to permanent structural modification (see Supplementary Note~1). 
By systematically mapping crystallographic orientations and $S_\mathrm{e}$, we demonstrate that ion-track formation in this low-symmetry monoclinic oxide is intrinsically anisotropic, challenging the conventional assumption of radial symmetry and revealing orientation-dependent thresholds, morphologies, and recovery kinetics.

\section*{Results} 

\subsection*{Energy deposition and initial damage production}

The multiscale simulation of SHI irradiation begins with Monte Carlo (MC) modeling of the initial ionization processes.
\verb|Geant4| software~\cite{agostinelli2003geant4} was utilized to simulate the resulting spatial distribution of electron excitations within the $\beta$-\ce{Ga2O3} target. 
The type and energy of incident ions are determined firstly to cover a wide range of $S_\mathrm{e}$.
In this work, four representative types of ions and the corresponding $S_\mathrm{e}$ are simulated, as listed in Table~\ref{tab:SHI}.
We note that the simulations of the O, Kr and Ta ion tracks align well with the experimental studies on SHI irradiation effects in $\beta$-\ce{Ga2O3} available in the literature~\cite{iancu2025intrinsic, ai2019radiation}.

\begin{table}[htbp!]
\begin{threeparttable}
\caption{Details of SHI irradiation ion types, ion energies, and corresponding calculated $S_\mathrm{e}$ values.}
 \label{tab:SHI}
\begin{ruledtabular}
\begin{tabular}{l c c}
Ion type        & Ion energy (MeV)      & $S_\mathrm{e}$ (keV/nm)       \\   
\colrule                                                              
 O~\cite{iancu2025intrinsic}  & 12.0                  & 3.02                          \\
 Xe             & 19.6                  & 10.00                         \\
 Kr~\cite{ai2019radiation}             & 460                   & 18.23                         \\
 Ta~\cite{ai2019radiation}             & 1179                  & 43.87                         \\
\end{tabular}
\end{ruledtabular}
\end{threeparttable}
\end{table}

\begin{figure*}[htbp!]
  \centering
  \includegraphics[width=\linewidth]{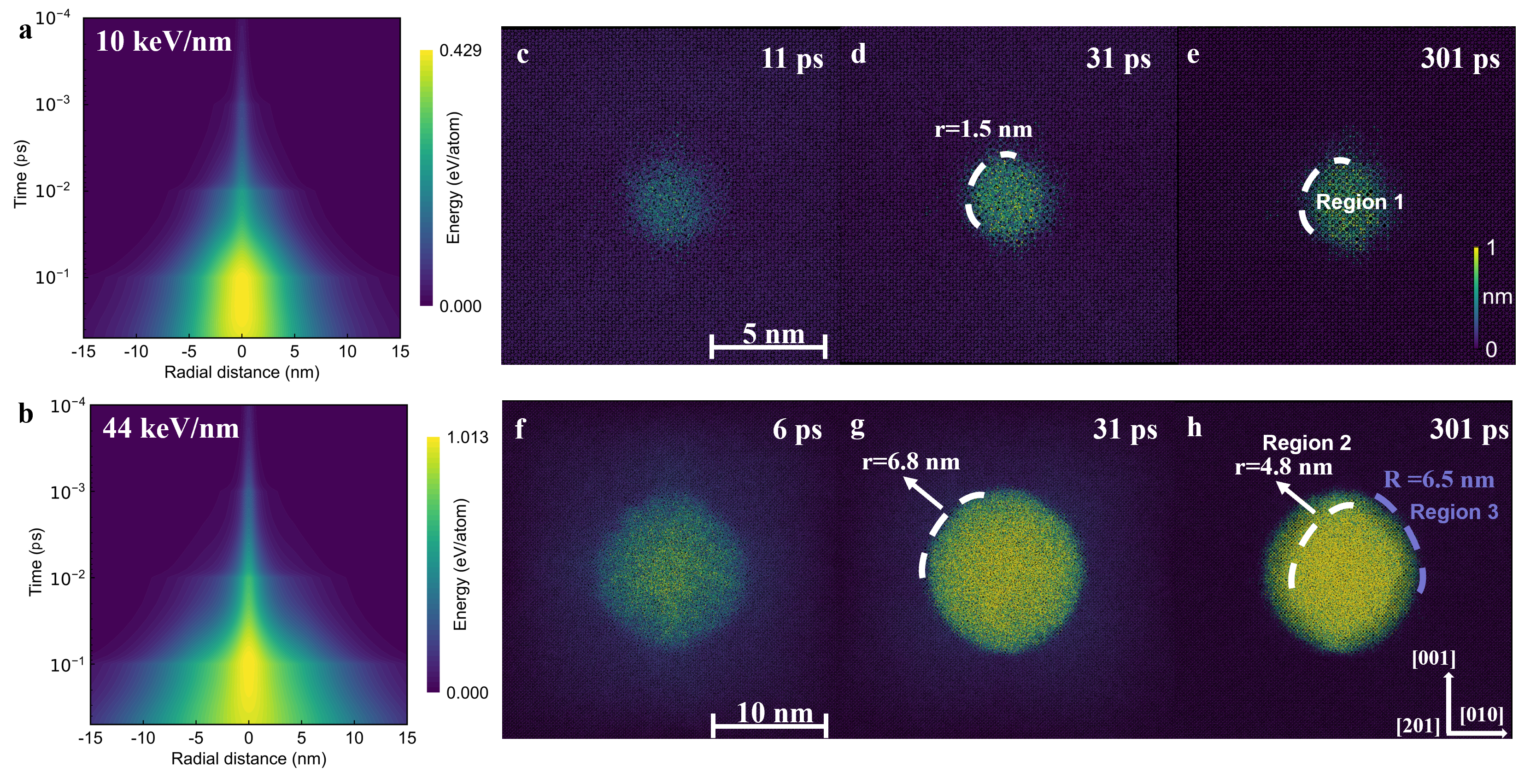}
  \caption{\textbf{Spatiotemporal lattice energy evolution and ion-track morphology under low and high \textit{S}\textsubscript{e}.}
  \textbf{a, b} Spatiotemporal profiles of lattice energy obtained from TTM simulations at $S_\mathrm{e}$ of 10~keV/nm and 44~keV/nm, respectively.
  \textbf{c-h} Cross-sectional morphologies from MD simulations with $S_\mathrm{e}=10$~keV/nm and 44~keV/nm, colored by atomic displacement magnitude.
  All images correspond to the central cross section of the simulated region, representing an effective sample thickness of 5~nm. 
  The view direction is perpendicular to the \hkl(100) plane.
  }
  \label{fig:displacement-Se}
\end{figure*}

Following the MC simulation of an SHI interaction with the electronic subsystem, the obtained distribution of the electronic energy loss was used as input in the analytical two-temperature model (TTM).
Within the TTM framework, the electronic and lattice subsystems are treated as two coupled heat baths.
Fig.~\ref{fig:displacement-Se}a, b illustrate the spatiotemporal evolution of energy within the lattice subsystem following a SHI irradiation with the $S_\mathrm{e}$ values of 10~keV/nm and 44~keV/nm, respectively. 
In both cases, the energy profiles in the lattice subsystem exhibit an initial sharp rise followed by a slow decay and spatial broadening.
This behavior reflects the rapid transfer of energy from the excited electrons to the lattice via the electron-phonon coupling and its subsequent redistribution in the lattice via thermal diffusion. 
Soon after the energy in the lattice reached its peak, the electronic and lattice subsystems reached a local thermal equilibrium ($T_\mathrm{e} \simeq T_\mathrm{l}$).
Under this condition, the distribution of the kinetic energies of the atoms in the lattice assumes a Gaussian-like shape. 
In these calculations, we observe that the total energy deposited to the lattice by a SHI increases monotonically with increasing $S_\mathrm{e}$. 
Moreover, higher $S_\mathrm{e}$ values produce higher energy peaks in the lattice and more extended high-energy regions, implying stronger local atomic displacements and an increased likelihood of thermally driven defect formation. 

In Fig.~\ref{fig:displacement-Se}c--h, we present the cross-sectional views of the resulting ion tracks obtained from ML-MD simulations, where the above energy deposition profiles are used as input to drive the atomic evolution.
For $S_\mathrm{e}=10$~keV/nm (Figs.~\ref{fig:displacement-Se}c--e) and $S_\mathrm{e}=44$~keV/nm (Figs.~\ref{fig:ion-track}f--h), the SHI tracks are shown for irradiation perpendicular to the \hkl(100) plane.
For the lower $S_\mathrm{e}$ of 10~keV/nm, we see only a small and highly localized region of atomic displacements that is formed at approximately 11~ps after ion impact (Fig.~\ref{fig:displacement-Se}c).
As the displaced atoms dissipate their residual kinetic energy, a transient lattice disorder emerges at around 30~ps, producing a short-lived amorphous pocket with a radius of roughly 1.5~nm. 
However, due to the limited energy deposition, this amorphous zone subsequently undergoes rapid structural recovery.
By the end of the simulation (301~ps), no stable amorphous domains persist, confirming that the deposited energy is insufficient to form a stable ion track or to induce permanent amorphization in $\beta$-\ce{Ga2O3}.

In contrast, for the high $S_\mathrm{e}$ of 44~keV/nm shown in Fig.~\ref{fig:displacement-Se}d, an extensive and severe disordering process begins, which progresses through several distinct stages. 
A significant atomic displacement is already evident at 6~ps after a SHI impact due to the substantially larger kinetic energy imparted to the lattice.
Subsequently, this initial damage zone expands radially, reaching a lateral extent of approximately 7~nm by 31~ps.
Such dynamic disordering is characteristic of ion-induced amorphization, and the central region of the ion track evolves into a fully amorphous structure where the energy deposition density is highest.
Notably, this amorphization process is transient. 
After approximately 31~ps, the tendency toward amorphization is markedly suppressed and gives way to a recrystallization process that dominates the subsequent structural evolution.
As illustrated by the time-sequential atomic configurations in Fig.~\ref{fig:displacement-Se}h, lattice fringes reappear in the peripheral region of the track (Region 3), indicating the nucleation and growth of recrystallized domains.
Ultimately, although a large fraction of atoms return to their original crystalline positions, a residual disordered core (Region 2) with a radius of 4.8~nm persists at the track center.

With respect to the other $S_\mathrm{e}$ values, an analogous structural evolution is observed, as shown in the Supplementary Fig.~S2. 
Under the very low $S_\mathrm{e}$ of 3~keV/nm, corresponding to irradiation by O ions, the overall crystal structure remains essentially intact, exhibiting negligible disruption after SHI irradiation.
Only minor lattice vibrations and local bond-angle fluctuations occur.
This observation is consistent with a recent experimental finding that no detectable ion tracks form in $\beta$-\ce{Ga2O3} under 12~MeV O ion irradiation~\cite{iancu2025intrinsic}.
The Wigner-Seitz defect analysis also confirms that no stable defects are generated under this condition.
For an intermediate $S_\mathrm{e}$ of 18~keV/nm, enhanced electronic energy deposition results in the formation of a transient amorphous region with the initial radius of approximately 6--8~nm immediately after the SHI impact.
In contrast to the more persistent amorphous regions produced at 44~keV/nm, the amorphous zone formed in this intermediate $S_\mathrm{e}$ regime exhibits partial recrystallization within 15--20~ps.
The spatial extent of the amorphous region becomes nearly equal to that of the recrystallized area.
This suggests that the competition between defect generation and thermal recovery reaches an approximate equilibrium under this excitation condition.

\subsection*{Atomic scale structural evolution}

\begin{figure*}[htbp!]
 \includegraphics[width=\linewidth]{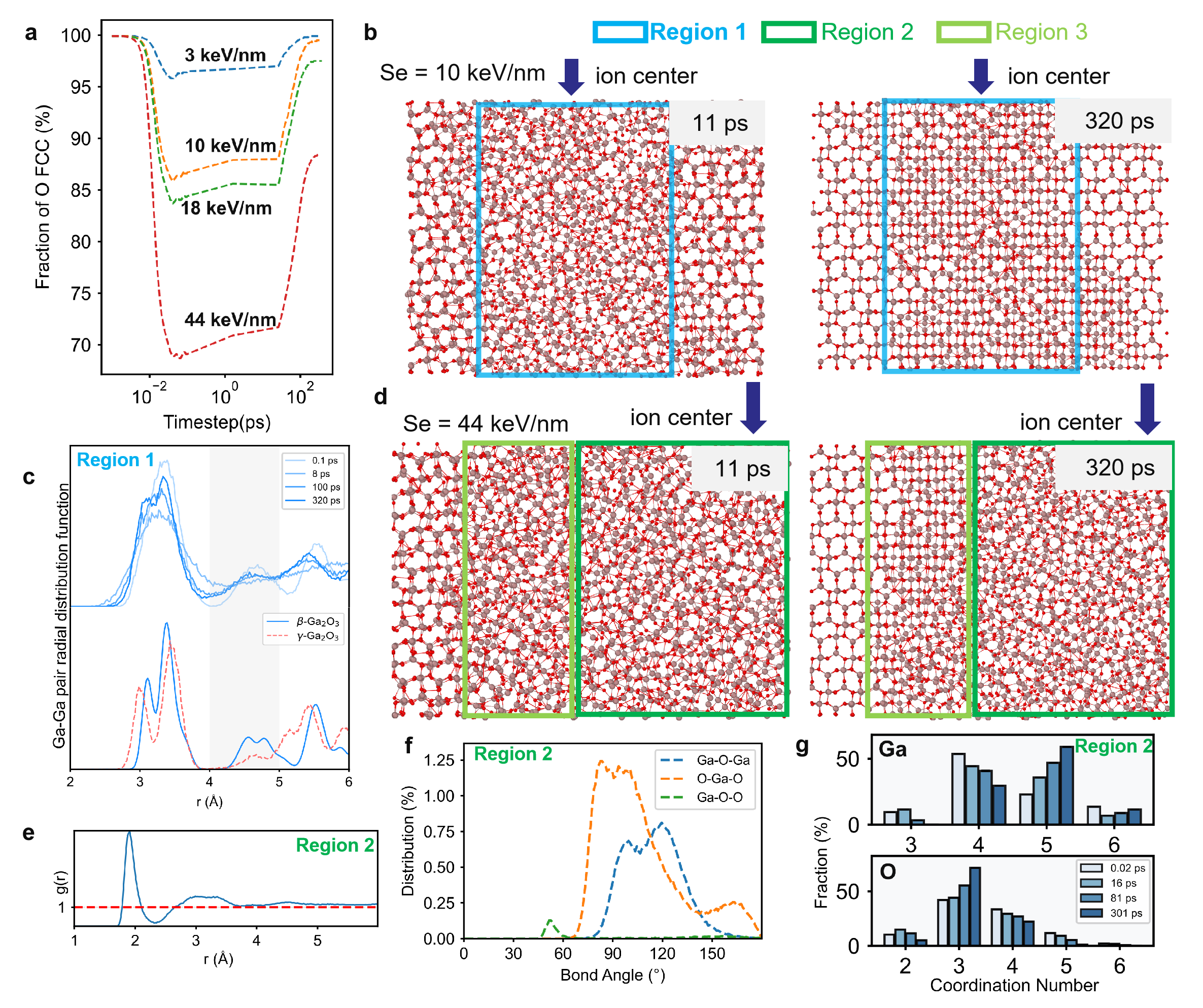}
 \caption{\textbf{Structural analysis of \textit{\textbeta}-\ce{Ga2O3} under SHI irradiation perpendicular to (1\,0\,0) plane at different electronic energy losses.} 
 \textbf{a} O FCC order parameter evolution at different $S_\mathrm{e}$ values.
 \textbf{b} Atomic configurations at 11~ps and 320~ps ($S_\mathrm{e} = 10$~keV/nm). 
 Brown and red balls are Ga and O atoms, respectively.
 \textbf{c} Evolution of Ga–Ga RDF of the track core at $S_\mathrm{e}$ of 10~keV/nm. (Region~1, $r \leq 2.5 $~nm).
 \textbf{d} Atomic configurations at 11~ps and 320~ps ($S_\mathrm{e} = 44$~keV/nm).
 \textbf{e} The overall RDF of the track core at $S_\mathrm{e}$ of 44~keV/nm (Region~2, $r \leq 4.5$~nm).
  \textbf{f} Bond-angle distribution of Region 2 at 320~ps. 
  \textbf{g} Coordination-number distributions of Region 2 (cutoff distance = 2.45 \r A).} 
 \label{fig:OFCC&RDF}
\end{figure*}

To elucidate the mechanism of damage formation in $\beta$-\ce{Ga2O3} under SHI irradiation, we performed a detailed structural analysis focusing on the entire evolution process.
Although Ga and O occupy multiple nonequivalent Wyckoff sites, the O sublattice preserves a face-centered-cubic (FCC) topology that serves as a robust order parameter for tracking crystalline recovery~\cite{zhao2025crystallization}. 
As shown in Fig.~\ref{fig:OFCC&RDF}a, the FCC fraction of O sublattice exhibits a universal two-stage behavior across all $S_\mathrm{e}$ values.
Rapid disorder within 0.1~ps, followed by gradual recovery extending to tens or even hundreds of picoseconds.
At low $S_\mathrm{e}$ (3~keV/nm), transient disordering is mild, and the O sublattice eventually returns to its pristine FCC arrangement.
No residual damage is detected, consistent with experimental reports showing the absence of ion tracks in $\beta$-\ce{Ga2O3} irradiated with low-$S_\mathrm{e}$ O ions~\cite{iancu2025intrinsic}.

At an intermediate $S_\mathrm{e}$ of 10~keV/nm, the O sublattice largely restores its FCC-like order, while the Ga sublattice exhibits only partial structural recovery.
As shown in Fig.~\ref{fig:OFCC&RDF}b, the atomic configuration undergoes substantial rearrangement during rapid thermal quench, but the residual disorder remains mainly associated with Ga sites.
The Wigner-Seitz defect analysis further indicates that only $\sim$2\% of the atoms form persistent defects, while these defects are overwhelmingly Ga vacancies ($\sim$98.5\%). 
This asymmetric recovery between the O and Ga sublattices drives the system into nonequilibrium reconstruction pathways, yielding a metastable $\gamma$-like configuration rather than restoring the pristine $\beta$-phase. 
This behavior is also clearly reflected in the Ga-Ga pair radial distribution functions (RDFs) in the core region (Region 1). 
As shown in Fig.~\ref{fig:OFCC&RDF}c, the RDF sharpens markedly, indicating the occurrence of recrystallization. 
However, the resulting peak positions and relative intensities do not match those of monoclinic $\beta$-\ce{Ga2O3}.
In particular, the absence of the characteristic 4--5~\r A peak of the second Ga-Ga shell confirms that the recovered structure does not revert to its original $\beta$ lattice.
Instead, the features of RDF are consistent with a metastable configuration of $\gamma$ phase, resembling structural transformations previously observed under high-dose, low-energy irradiation dominated by nuclear stopping~\cite{zhao2025crystallization}.
In contrast, the present transition arises solely from electronic energy deposition, demonstrating that electronic excitations alone are sufficient to induce and stabilize the $\gamma$-like phase.
The resulting $\gamma$-phase interface remains stable throughout the simulation, highlighting its robust metastable persistence.

In the higher $S_\mathrm{e}$ regime, the system exhibits a qualitatively different structural response.
At the highest $S_\mathrm{e}$ of 44~keV/nm, the ion track no longer remains structurally uniform but instead displays pronounced radial heterogeneity.
Depending on the local structural characteristics, the track can be divided into three distinct radial regions: a central core (Region~2), an intermediate shell (Region~3), and the surrounding crystalline matrix, forming a three-phase core-shell morphology, as illustrated in Fig.~\ref{fig:OFCC&RDF}d. 
To identify the structural nature of the track core, we firstly examine the RDF and the bond-angle distribution of Region~2. 
As shown in Fig.~\ref{fig:OFCC&RDF}e, the total RDF rapidly converges to unity beyond the first-neighbor shell, indicating the absence of long-range order.
The bond-angle distribution also reveals broad distributions of Ga-O-Ga and O-Ga-O triplet angles ranging from 60$\degree$ to 130$\degree$.
In particular, the O-Ga-O angles exhibit characteristic peaks near 83$\degree$, 94$\degree$, and 163$\degree$, while the Ga-O-Ga angles preferentially occur around 100$\degree$ and 120$\degree$.
These angular features closely resemble those of amorphous structures obtained from direct annealing simulations~\cite{jiahui-amorphous2025large}, confirming the amorphous nature of the track core.
In addition, Ga-O-O triplets with a characteristic angle near 52$\degree$ are identified, providing a structural origin for the small peak previously reported in Ref.~\cite{zhao2023complex}.

The structural evolution of the amorphous core is further quantified by the coordination-number distributions of Ga and O atoms, as shown in Fig.~\ref{fig:OFCC&RDF}g.
In the pristine $\beta$ phase, Ga atoms predominantly exhibit fourfold and sixfold coordination with the ratio of $1{:}1$.
Upon SHI irradiation, the fraction of sixfold Ga atoms decreases sharply from $\sim$50\% to $\sim$10\% within the first 0.1~ps and remains at this low level thereafter.
Meanwhile, the population of fourfold Ga atoms initially increases, reflecting a rapid reduction in coordination.
As the system evolves, these fourfold Ga atoms progressively transform into fivefold configurations, which ultimately dominate the amorphous core, reaching nearly 60\% in the final state.
Throughout the first 80~ps, a small fraction ($\sim$10\%) of threefold Ga atoms is observed, but this coordination state is transient and disappears in the final configuration.
In general, the evolution of the Ga coordination indicates a sequential transformation from six to four and finally to fivefold states, consistent with the preferential displacement of sixfold Ga atoms reported in threshold displacement energy calculations~\cite{he2024threshold}.
In contrast, the oxygen sublattice exhibits a distinct coordination evolution.
Initially, threefold and fourfold O atoms are present in the ratio of $2{:}1$.
Both populations decrease rapidly within the first 0.1~ps after irradiation, resulting in twofold and fivefold states.
As the amorphous structure stabilizes, the fraction of threefold O atoms increases steadily, eventually reaching 71.5\% in the final state, while the population of fourfold O stabilizes at approximately 22\%.
Higher coordination states remain rare, with only $\sim$5\% of O atoms being twofold and fewer than 1\% exhibiting fivefold or sixfold.
These results indicate that, despite substantial disordering, the O sublattice within the ion-track core retains a predominantly low-coordination character.
In addition, the RDFs of the final intermediate shell (Region~3) closely resemble those observed in the core of the track formed at a lower $S_\mathrm{e}$ of 10~keV/nm (Region~1). 
Accordingly, the ion track induced by the highest $S_\mathrm{e}$ can be inferred to consist of three structurally distinct forms of \ce{Ga2O3}: an amorphous core (Region~2), a $\gamma$-phase shell (Region~3), and the surrounding pristine $\beta$ matrix.

\begin{figure*}[htbp!]
 \includegraphics[width=\linewidth]{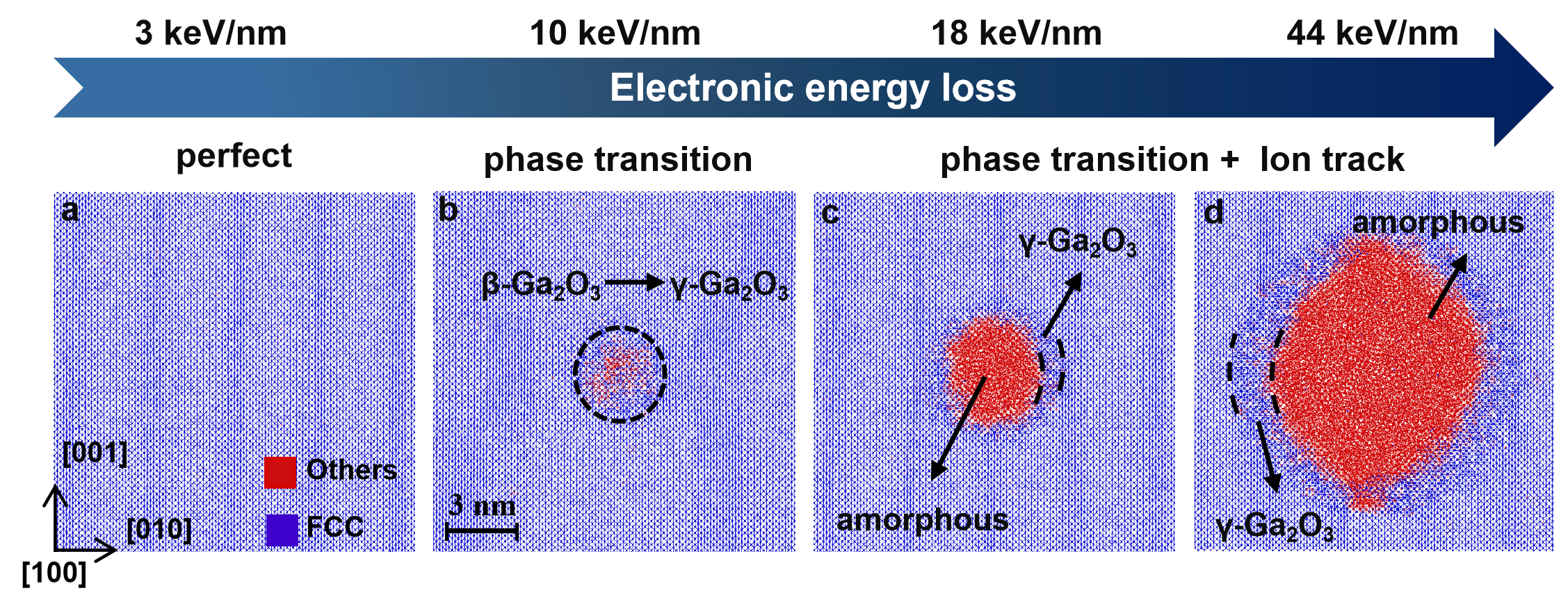}
 \caption{\textbf{Final state of oxygen lattice order mapping under different SHI irradiation perpendicular to (1\,0\,0) plane.}
  \textbf{a} 3~keV/nm: perfect oxygen lattice order, no phase transition or ion track. 
    \textbf{b} 10~keV/nm: phase transition from $\beta$-Ga$_2$O$_3$ to $\gamma$-Ga$_2$O$_3$ without ion track. 
    \textbf{c} 18~keV/nm and \textbf{d} 44~keV/nm: phase transition ($\beta$$\rightarrow$$\gamma$) accompanied by an ion track. }
 \label{fig:ion-track}
\end{figure*}

Building on the analysis described above of the local structural responses, we next examine how these atomic-scale transformations manifest at the mesoscale in the form of ion-track morphology.
Fig.~\ref{fig:ion-track} shows the final-state oxygen lattice-order maps for SHI irradiation perpendicular to the \hkl(100) plane with increasing $S_\mathrm{e}$.
A clear sequence of distinct structural responses is identified as $S_\mathrm{e}$ increases. 
At low $S_\mathrm{e}$ (3~keV/nm), the transient lattice disorder induced by electronic excitation fully recovers, leaving the crystalline framework essentially intact. 
At intermediate $S_\mathrm{e}$ (10 keV/nm), partial melting followed by recrystallization gives rise to a metastable $\gamma$-\ce{Ga2O3} phase within the irradiated region. 
At high $S_\mathrm{e}$ (18 keV/nm and 44 keV/nm), a well-defined core-shell ion-track morphology emerges, consisting of a highly disordered or amorphous core surrounded by a recrystallized shell dominated by the $\gamma$ phase. 
This progression underscores the strong dependence of SHI-induced structural evolution on the deposited electronic energy density.
Notably, the spatial distribution of FCC O atoms becomes strongly anisotropic at the highest $S_\mathrm{e}$, suggesting that the resulting ion-track morphology is highly sensitive to crystallographic orientation. 

\subsection*{Anisotropic Track Morphology}

\begin{figure*}[htbp!]
  \centering
  \includegraphics[width=16cm]{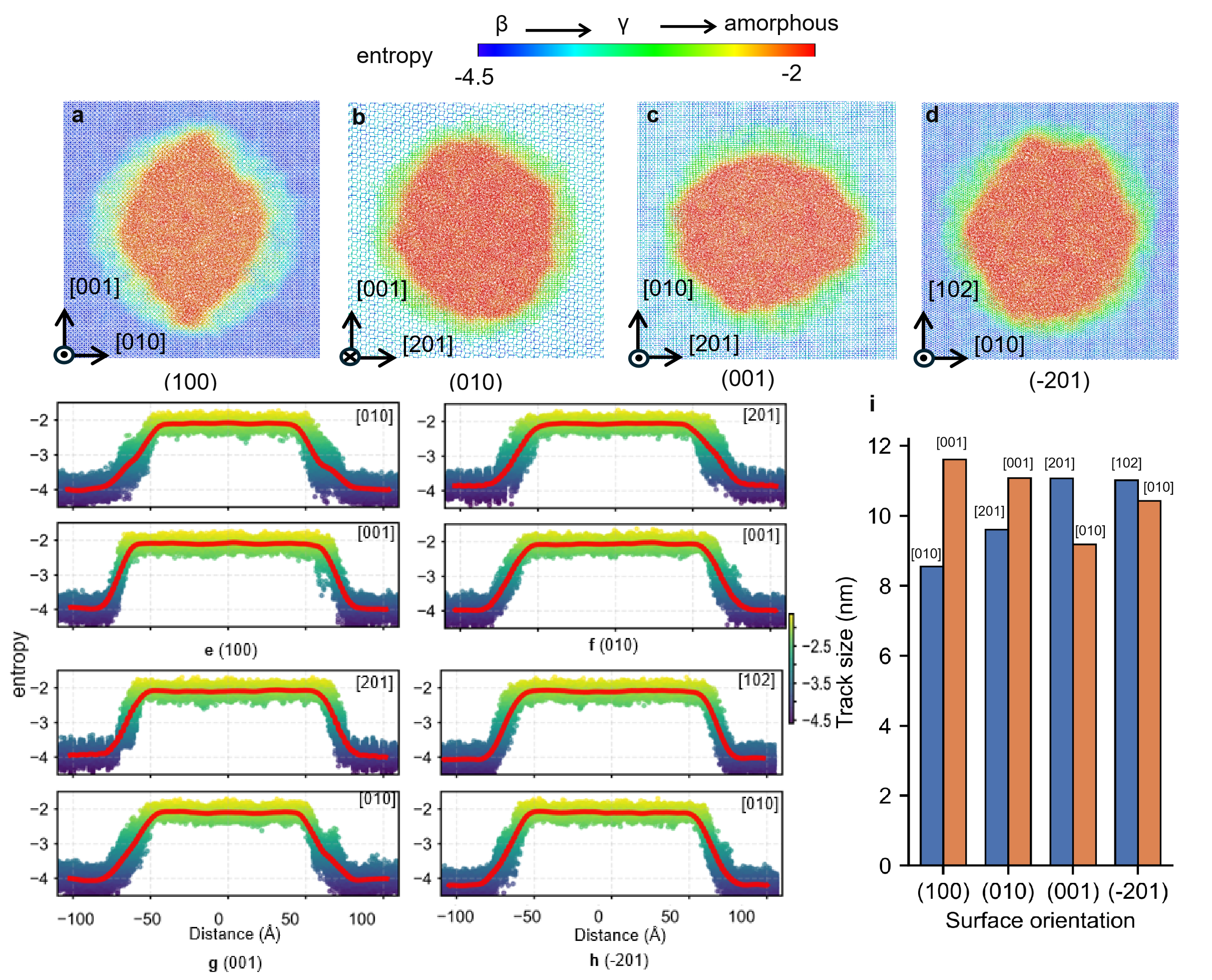}
\caption{\textbf{Ion-track morphology characterized by local structural entropy in \textit{\textbeta}-\ce{Ga2O3}}.
 \textbf{a--d} Spatial maps of the local structural entropy for ion tracks formed under SHI irradiation perpendicular to the \hkl(100), \hkl(010), \hkl(001), and \hkl(-201) crystallographic planes at $S_\mathrm{e}$ of 44~keV/nm.
 \textbf{e--h} One-dimensional entropy profiles extracted along the lateral principal crystallographic axes, averaged within a 1.5~nm wide slab, as a function of the radial distance from the track center.
 The red line denotes the mean entropy at each distance.
  \textbf{i} Ion track size under SHI irradiation perpendicular to the \hkl(100), \hkl(010), \hkl(001), and \hkl(-201) crystallographic planes at $S_\mathrm{e}$ of 44~keV/nm.
 All analyses are performed within a 5~nm thick region centered on the simulation cell.
}
  \label{fig:entropy}
\end{figure*}

To go beyond qualitative visualization, we employ the local configurational entropy ($\Omega$)~\cite{ovitostukowski2009visualization, SFliu2026anisotropic} as a mesoscale structural descriptor to 
quantify the ion-track morphologies across four representative crystallographic planes: \hkl(100), \hkl(010), \hkl(001), and \hkl(-201).
As shown in Figs.~\ref{fig:entropy}a--d, the ion-track morphology exhibits a striking dependence on the irradiation direction at the highest $S_\mathrm{e}$ of 44~keV/nm.
For SHI irradiation perpendicular to \hkl(100) plane, the track cross section exhibits an elongated elliptical, or nearly rhombic shape. 
In contrast, irradiation perpendicular to \hkl(010) plane results in a distinctly asymmetric track profile, reflecting the strongly anisotropic atomic arrangement within the \hkl(010) plane. 
A comparable elliptical morphology is obtained for SHI irradiation perpendicular to \hkl(001) direction. 
Remarkably, when viewed the SHI irradiation perpendicular to the \hkl(-201) plane, the ion track adopts a quasi-hexagonal geometry, indicating enhanced structural coherence associated with this crystallographic orientation.

Due to the pronounced anisotropy of $\beta$-\ce{Ga2O3}, a unique determination of the ion-track size from atomistic configurations alone is nontrivial.
We therefore analyze one-dimensional profiles of the $\Omega$ along the lateral principal axes passing through the track center, as shown in Fig.~\ref{fig:entropy}e--h. 
Despite the strong projection-dependent track shapes, all profiles exhibit a common entropy topology: $\Omega \simeq -4$ in the crystalline matrix, a sharp increase at intermediate distances (5.0--7.5~nm) corresponding to a structurally transformed shell associated with a transient $\gamma$ phase, and a central plateau with $\Omega \in [-2.5,-2.0]$ defining the fully amorphous track core. 
This universal entropy further demonstrates that ion tracks in $\beta$-\ce{Ga2O3} prefer the multilayer core-shell structure, independent of crystallographic orientation.

Based on this entropy-based criterion, the size of ion-track size are summarized in Fig.~\ref{fig:entropy}i.
Local structural entropy of other $S_\mathrm{e}$ are displayed in Supplementary Figs.~S3 and S4.
Irrespective of the irradiation direction, the final amorphous core is consistently more confined along the \hkl[010] direction.
In particular, for SHI irradiation perpendicular to the \hkl(100) and \hkl(001) planes, recovery along \hkl[010] is significantly more efficient than along other directions, resulting in the flattened elliptical track morphologies shown in Fig.~\ref{fig:entropy}a, c. 
This anisotropic recovery behavior is in excellent agreement with the experimental observations~\cite{ion_tracy2016anisotropic}, who reported negligible lattice swelling along \hkl[010], in contrast to pronounced expansion along \hkl[100] and \hkl[001]. 
More broadly, recent studies~\cite{wierbik2026anisotropic} have demonstrated that the anisotropy of ion-track radii is primarily governed by the elastic properties of material. 
For highly anisotropic elastic tensor of $\beta$-\ce{Ga2O3}, our MD simulations also reveal an exceptionally large Young’s modulus along the \hkl[010] direction, which is approximately 284~GPa (see Supplementary Fig.~S5). 
The enhanced stiffness along \hkl[010] promotes more efficient stress relaxation for the atomic rearrangement, facilitating recrystallization and effectively suppressing ion-track expansion along this direction. 
Consequently, these results demonstrate a direct link between elastic anisotropy and anisotropic ion-track recovery in $\beta$-\ce{Ga2O3}, establishing direction-dependent elastic stiffness as a key intrinsic factor governing ion-track morphology in low-symmetry oxides.

With respect to the ion-track size, the residual ion-track size increases systematically as the $S_\mathrm{e}$ increases, as displayed in Supplementary Table~S1
At $S_\mathrm{e}$ of 18~keV/nm, the simulated ion-track diameters remain limited to 2--4.2~nm, accompanied by a high recovery ratio approaching $\sim$47\% to $75\%$. 
In contrast, at higher $S_\mathrm{e}$ ($44$~keV/nm), the recovery efficiency drops to 26\%--44\%, resulting in a pronounced expansion of the amorphous core to diameters of 8.55--11.61~nm. 
The residual ion-track size of \hkl(100) plane remains comparatively smaller than that of other orientations at 18~keV/nm and 44~keV/nm.
This behavior reflects a superior recrystallization capability of the \hkl(100) orientation under intense SHI irradiation. 
In contrast, the \hkl(-201) plane, characterized by a significantly reduced recovery efficiency, accumulates more severe residual damage and develops larger amorphous cores under comparable irradiation conditions. 
Additionally, the simulated ion-track diameters for the \hkl(100) orientation are in quantitative agreement with available experimental measurements over a comparable range of electronic energy losses.
At highest $S_\mathrm{e}$ (44~keV/nm), our simulations yield a track diameter of 8.55~nm and 11.61~nm, which is consistent with the value of $7.8 \pm 0.9$~nm reported by Ai \textit{et al.}~\cite{ai2019radiation} at 41~keV/nm and $8.3 \pm 0.4$~nm measured by Tracy \textit{et al.}~\cite{ion_tracy2016anisotropic} for 946~MeV Au-ion irradiation ($S_\mathrm{e} \simeq 45$~keV/nm). 
At a lower $S_\mathrm{e}$ of 18.3~keV/nm, Ai \textit{et al.} observed a significantly reduced track diameter of $2.2 \pm 0.4$~nm, which is likewise consistent with the pronounced shrinkage of the track core obtained in our simulations (1.97~nm and 3.12~nm). 
This level of agreement across different $S_\mathrm{e}$ regimes confirms that the present simulations reliably capture both the characteristic length scale of ion-track formation and the orientation-dependent recovery behavior in $\beta$-\ce{Ga2O3}.

In contrast to the high $S_\mathrm{e}$ regime, the radiation response at low electronic stopping power ($S_\mathrm{e}=10$~keV/nm) is governed by a fundamentally different mechanism. As summarized in Supplementary Table~S1, only the \hkl(100) plane exhibits a detectable recovered $\gamma$ phase, whereas the other orientations display merely sparse point defects.
This indicates that the \hkl(100) plane is intrinsically more susceptible to SHI irradiation under low-$S_\mathrm{e}$ conditions.
To elucidate this different response, we quantified the fraction of atoms with kinetic energy exceeding $0.5$~eV before the atomic response.
Although an identical lattice energy profile was imposed for all orientations, the \hkl(100) plane consistently exhibits the highest fraction of high $E_\mathrm{k}$ atoms (see Supplementary Table~S1). We attribute it to the significant difference of atomic planar density. The atomic planar density of \hkl(100) plane is $\sim$0.28~atom/\r A$^2$, which is significantly higher than those of the \hkl(010), \hkl(001), and \hkl(-201) planes (approximately 0.13, 0.14, and 0.25~atom/\r A$^2$, respectively). Therefore, more atoms, accounting for the enhanced radiation response of the \hkl(100) plane at low $S_\mathrm{e}$. 

In addition, though a pronounced crystallographic anisotropy is observed in the final ion-track morphology, the combined size of the amorphous core and the surrounding $\gamma$-phase region remains nearly constant for all crystallographic orientations, with characteristic radii of approximately 7.5~nm and 15~nm at 18~keV/nm and 44~keV/nm, respectively. 
This invariance demonstrates that the initial radial distribution of electronic energy deposition and the ensuing amorphization process are largely isotropic. 
Consequently, the pronounced anisotropy in the final amorphous core size must originate predominantly from orientation-dependent post-irradiation recovery processes rather than from the primary damage stage.

In summary, the entropy-based analysis reveals a clear separation between the primary damage stage and the post-irradiation recovery process in SHI-irradiated $\beta$-\ce{Ga2O3}.
While the initial electronic energy deposition and amorphization are essentially isotropic, the final ion-track morphology is governed by strongly orientation-dependent recovery dynamics.
The pronounced confinement of the amorphous core along \hkl[010] originates from the exceptionally high elastic stiffness along this direction, which promotes efficient stress relaxation and recrystallization during the thermal-spike cooling stage.
At high energy losses, this recovery anisotropy dictates the residual track size, whereas at low energy losses, the radiation response is controlled by the competition between deposited energy and crystallographic atomic planar density.
These findings establish elastic anisotropy as a key factor controlling ion-track recovery in low-symmetry oxides and provide a unified framework for understanding orientation-dependent SHI effects in $\beta$-\ce{Ga2O3}.
 
\section*{Methods}

\textbf{Monte Carlo (MC) Simulation.} The interaction of SHIs with \ce{Ga2O3} was simulated using the MC particle transport code \verb|Geant4|.
A rectangular \ce{Ga2O3} target was constructed with a top surface area of $39.9 \times 39.9$~nm$^{2}$.
Heavy ions were injected vertically along the center of the top surface.
The physical processes governing heavy-ion interactions were described using the standard \verb|QGSP_BIC| physics list.
To enhance the accuracy of low-energy electromagnetic interactions and elastic ion-atom collisions, the ``\verb|G4EmStandardPhysics_option4|'' module was additionally incorporated.
This combination allows for a reliable description of both electronic stopping loss and nuclear stopping loss.
The total ionizing energy deposition was accumulated independently in each cell.
After simulating a statistically significant number of ion impacts, the average spatial distributions of ionizing energy deposition were obtained.

\textbf{Two temperature model (TTM).} The TTM equations were solved numerically by our self-developed TTM code, \verb|Femto-2TM|~\cite{liang2026femtosecond}, yielding the spatiotemporal distribution of the energy transferred to the atomic lattice.
The target was discretized into a three-dimensional grid of $133 \times 133 \times 9$ cells within $39.9 \times 39.9 \times 9~\text{nm} $~nm$^{3}$ . The numerical equations are given by:
\begin{equation}
\left\{
\begin{aligned}
C_\mathrm{e}(T_\mathrm{e}) \frac{\partial T_\mathrm{e}}{\partial t}
&= \nabla \cdot \bigl[ K_\mathrm{e}(T_\mathrm{e}) \, \nabla T_\mathrm{e} \bigr]
   - g \left( T_\mathrm{e} - T_\mathrm{l} \right)
   + A_\mathrm{e}(\mathbf{r}, t), \\
C_\mathrm{l}(T_\mathrm{l}) \frac{\partial T_\mathrm{l}}{\partial t}
&= \nabla \cdot \bigl[ K_\mathrm{l} \, \nabla T_\mathrm{l} \bigr]
   + g \left( T_\mathrm{e} - T_\mathrm{l} \right),
\end{aligned}
\right.
\end{equation}
where $T_\mathrm{e}$ and $T_\mathrm{l}$ are the electron and lattice temperatures, respectively. $C_\mathrm{e}$ and $C_\mathrm{l}$ denote the corresponding heat capacities. $K_\mathrm{e}$, $K_\mathrm{l}$ are the thermal conductivities and $g$ is the electron-phonon coupling constant. $A_\mathrm{e}(\mathbf{r},t)$ represents the electronic energy source term derived from the ionizing energy deposition calculated by MC simulations. Details of the parameter can be found in the supporting information.

\textbf{Molecular dynamics (MD) simulation.} All MD simulation were employed by \verb|LAMMPS| software~\cite{lammps}. 
The energy deposition was modeled by imparting initial kinetic energy from TTM simulation.
Supercells on the order of $150 \times 300 \times 300$~\r A$^{3}$ containing about 1152,000 atoms, were constructed for SHI radiation perpendicular to the \hkl(100) plane for example. 
The simulation system size of other planes are similar.
The simulated supercell was first equilibrated in isothermal-isobaric (\textit{NPT}) ensemble at 300~K and 0~bar. 
Subsequently, the SHI irradiation process was simulated in the microcanonical (\textit{NVE}) ensemble, while a 300~K thermostatted outer layer was applied to act as a heat sink.
The Langevin thermostat was applied to boundary regions parallel to the SHI trajectory with a layer thickness of 5~\r A. 
Periodic boundary conditions were applied in all three dimensions. 
All MD simulations were conducted by the fast machine-learned tabGAP potential~\cite{zhao2023complex} which has been validated for simulating complex radiation-induced phenomena relevant to this study~\cite{he2024threshold, zhao2025crystallization}. 
A variable time step from 0.01~fs to 1~fs was used and a total of 320~ps is simulated. 
Three independent runs were performed for each configuration to ensure statistical significance.
Structural changes during the simulations were characterized using the \verb|OVITO| software~\cite{ovitostukowski2009visualization}.

\section*{Data availability}

The \ce{Ga2O3} ML-IAP parameter files used to run classical MD simulations are openly available at \url{https://doi.org/10.6084/m9.figshare.21731426}.  
The raw data of TTM simulations of SHI and the MD simulation movie published in this paper are openly available at \url{https://doi.org/10.6084/m9.figshare.32504067}.  
Further data are available from the corresponding author upon reasonable request.

\section*{Code availability}

The MC, MD, and visualization software used in this work are \verb|Geant4|, \verb|LAMMPS| and \verb|OVITO|, respectively, which are openly available online from the corresponding developers and maintainers.
The TTM code is \verb|Femto-2TM|, which is available from the corresponding author upon reasonable request.

\bibliography{main}

\clearpage

 \section*{Acknowledgements}

This work is funded by Innovative Scientific Program of CNNC, the National Natural Science Foundation of China (NSFC) (No. 12405298) and China Postdoctoral Science Foundation funded project (No. 2024M762610). 

\section*{Author Contributions}

H.H. and J.L.Z. conceived the idea and designed the simulations. 
J.Y.L. and S.W.H. developed the \verb|Femto-2TM| code and performed the simulations.  
J.H.Z., Z.Q.C., and T.S. performed the simulations and collected the data.
Y.W.Z., F.D., and H.Z. contributed to the data analysis and validation. 
H.H. and J.L.Z. wrote the initial draft of the manuscript. 
C.H.H., H.Z., F.D., and J.L.Z. supervised the project, secured funding, and thoroughly revised the manuscript. 
All authors discussed the results, provided critical feedback, and approved the final manuscript.

\section*{Ethics declarations}

\subsection*{Competing interests}

The authors declare that they have no financial or personal relationships with individuals or organizations that could potentially influence the work reported in this paper. 

\section*{Supplementary information}

Supplementary Notes 1--4.

\end{document}